\documentclass[12pt,english]{article}

\usepackage[]{natbib}
\usepackage{url}
\makeatletter
\def\url@leostyle{%
    \def\UrlFont{\sf}}{\def\UrlFont{\small\ttfamily}}
\makeatother
\usepackage{geometry}
\usepackage[font=small,labelfont=bf]{caption}

\usepackage[hidelinks,breaklinks]{hyperref}
\usepackage[hyphenbreaks]{breakurl}

\usepackage{authblk}

\usepackage{amsmath}

\usepackage[T1]{fontenc}
\usepackage[charter]{mathdesign} 

\usepackage{microtype}

\usepackage{graphicx}

\usepackage[english]{babel}

\numberwithin{equation}{section}

\newcommand{\documentTitle}{Information Causality, the Tsirelson Bound, and the
  `Being-Thus' of Things}

\title{\documentTitle\thanks{This is the submitted version of an article
    published in the journal: \emph{Studies in History and Philosophy of Modern
      Physics.} There may be changes made in the published version which
    are not reflected in this preprint. When citing this paper, \textbf{please
      refer to the published version}, which can be accessed via:
    \href{https://doi.org/10.1016/j.shpsb.2018.05.001}
         {doi.org/10.1016/j.shpsb.2018.05.001}.}%
  $\mbox{ }^,$\thanks{I am indebted to the late William Demopoulos for
  sharing a draft of the chapter ``Quantum Reality'' with me from his as of
  yet unpublished monograph \emph{On Theories}. The present paper is
  motivated, illuminated, and informed both by that chapter and by my
  conversations with Bill during the years 2013--2017. Thanks also to Jeffrey
  Bub, Laura Felline, Sona Ghosh, Markus M\"uller, and Ryan Samaroo for their
  helpful and constructive comments on a previous draft of this paper. And
  thanks to those in attendance for my presentations of this paper at the
  University of Geneva and at the Workshop in Memory of William Demopoulos at
  the University of Western Ontario in September 2017 for their helpful and
  constructive questions. This project was supported financially by the
  Foundational Questions Institute (FQXi) and by the Rotman Institute of
  Philosophy. I hope for but do not claim the endorsement of my conclusions
  by any persons or institutions mentioned above.}}

\author[a, b]{Michael E. Cuffaro}
\affil[a]{{\small Rotman Institute of Philosophy, University of Western
    Ontario}}
\affil[b]{{\small Munich Center for Mathematical Philosophy,
    Ludwig-Maximilians-Universit\"at M\"unchen}}
\date{}

\begin{document}

\maketitle

\pagestyle{plain}
\thispagestyle{empty}

\begin{abstract}
The principle of `information causality' can be used to derive an upper
bound---known as the `Tsirelson bound'---on the strength of quantum mechanical
correlations, and has been conjectured to be a foundational principle of
nature. To date, however, it has not been sufficiently motivated to play such a
foundational role. The motivations that have so far been given are, as I argue,
either unsatisfactorily vague or appeal to little if anything more than
intuition. Thus in this paper I consider whether some way might be found to
successfully motivate the principle. And I propose that a compelling way of so
doing is to understand it as a generalisation of Einstein's principle of the
mutually independent existence---the `being-thus'---of spatially distant
things. In particular I first describe an argument, due to Demopoulos, to the
effect that the so-called `no-signalling' condition can be viewed as a
generalisation of Einstein's principle that is appropriate for an irreducibly
statistical theory such as quantum mechanics. I then argue that a compelling
way to motivate information causality is to in turn consider it as a further
generalisation of the Einsteinian principle that is appropriate for a theory of
communication. I describe, however, some important conceptual obstacles that
must yet be overcome if the project of establishing information causality as a
foundational principle of nature is to succeed.
\end{abstract}

\section{Introduction}
\label{sec:intro}

Answering the question of precisely what distinguishes our experience with
quantum as opposed to classical physical phenomena has historically been a
central element of the overall project of interpreting quantum theory. For
\citet[]{schrodinger1935}, for instance, the sole distinguishing feature of
quantum theory was none other than entanglement, while for Feynman the one and
only quantum mystery was self-interference \citep[vol. 3,
  1-1]{feynman1964}. The question continues to occupy many. However in much of
the more recent literature it has taken on a different form. That is, it has
become one of specifying a set of appropriately motivated constraints or
`principles' that serve to distinguish quantum from classical
theory. \citet*[]{clifton2003}, for instance, prove a theorem which they argue
shows quantum mechanics to be essentially characterisable in terms of a small
number of information-theoretic constraints. \citet[]{spekkens2007}, meanwhile,
shows that features often thought of as distinctively quantum can be manifested
in a toy classical theory to which one adds a principled restriction on the
maximal obtainable knowledge of a system.\footnote{For a discussion of both
  \citeauthor[]{clifton2003}'s and \citeauthor[]{spekkens2007}' results, and
  of the project in general, see \citet[]{myrvold2010}; and see also
  \citet[]{felline2016}.}

One feature that quantum and classical theory have in common is that the
correlations manifested between the subsystems of a combined system satisfy the
condition that the marginal probabilities associated with local experiments on a
subsystem are independent of which particular experiments are performed on the
other subsystems. It is a consequence of this condition that it is impossible to
use either a classically correlated or entangled quantum system to signal faster
than light. For this reason the condition is referred to as the `no-signalling'
condition or principle, even though the condition is not a relativistic
constraint \emph{per se}.

Quantum and classical theory do not exhaust the conceivable ways in which the
world could be. The world could be such that neither quantum nor classical
theory are capable of adequately describing the correlations between subsystems
of combined systems. In particular the world could be such that correlations
\emph{stronger} than quantum correlations are possible within it. In a landmark
paper, \citet[]{popescu1994} asked the question of whether all such correlations
must violate the no-signalling condition. The surprising answer to this question
is no. As they showed, there do indeed exist conceivable correlations between
the subsystems of combined systems that are stronger than the strongest possible
quantum correlations---i.e. such that they exceed the so-called `Tsirelson
bound' \citep[]{tsirelson1980}---and yet non-signalling.

\citeauthor[]{popescu1994}'s result raises the question of whether some
motivated principle or principles can be given which would pick out quantum
theory---or at least some restricted subset of theories which includes quantum
theory---from among the space of conceivable non-signalling physical theories in
which correlations at or above the Tsirelson bound occur. This question has
developed into an active research program. A particularly important result
emerging from it is that of \citet[]{pawlowski2009}, who show that one can in
fact derive the Tsirelson bound from a principle they call `information
causality', which they describe as a generalisation of no-signalling applicable
to experimental setups in which the subsystems of a combined system
(e.g. spatially separated labs) may be subluminally communicating classical
information with one another. \citeauthor[]{pawlowski2009} conjecture that
information causality may be a foundational principle of nature.

Below I will argue that, suitably interpreted \citep[][]{bub2012}, the principle
can be regarded as a useful and illuminating answer to the question of what the
Tsirelson bound expresses about correlations which exceed it. However I will
argue that if one wishes to think of information causality as a fundamental
principle of nature---in the sense that theories which violate the principle
should thereby be regarded as unphysical or in some other sense
impossible---then it requires more in the way of motivation than has hitherto
been given.

What has typically been appealed to previously to motivate the principle is the
intuition that a world in which information causality is not satisfied would be
`too simple' \citep[p. 1101]{pawlowski2009}, or `too good to be true'
(\citealt[p. 180]{bub2012}, \citealt[p. 187]{bub2016}); that it would allow one
to ``implausibly'' access remote data \citep[ibid.]{pawlowski2009}, and that
``things like this should not happen'' \citep[p. 429]{pawlowski2016}. I will
argue below that these statements are unsatisfactorily vague. Nevertheless I
will argue that they gesture at something that is importantly right; although
they are right in, perhaps, a different sense than their authors envision.

More specifically, in contrast to \citet[]{bub2012}, who in his otherwise
illuminating analysis of information causality argues that it is misleadingly
characterised as a generalisation of the no-signalling principle, I will argue
that information causality can indeed be regarded as generalising no-signalling
in a sense. To clarify this sense I will draw on the work of
Demopoulos,\footnote{\label{fn:demo}I am referring to the chapter ``Quantum
  Reality'' of Demopoulos's monograph \nocite{demopoulosForth}\emph{On
    Theories}, which is currently being prepared for posthumous publication.}
who convincingly shows that no-signalling can itself be thought of as a
generalisation, appropriate for an irreducibly statistical theory such as
quantum mechanics, of Einstein's principle of the mutually independent
existence of spatially distant things. Einstein regarded this principle as
necessary for the very possibility of `physical thought', and argued that it is
violated by quantum mechanics \citep[p. 187]{howard1985}. However, suitably
generalised and interpreted as a constraint on physical practice, Demopoulos
convincingly argues that Einstein's principle is in that sense satisfied both
in Newtonian mechanics (despite its being an action-at-a-distance theory), and
indeed (somewhat ironically\footnote{Demopoulos's `judo-like' argumentative
  manoeuvre is reminiscent of Bell's \citep[cf.][p. 41]{shimony1984}.}) that
it is satisfied in quantum mechanics, wherein it is expressed by none other
than the no-signalling condition.

Coming back to information causality, I will then argue that it can likewise be
thought of as a further generalisation of Einstein's principle that is
appropriate for a theory of communication. As I will clarify, in the context of
the experimental setups to which the principle is applicable, a failure of
information causality would imply an ambiguity in the way one distinguishes
conceptually between the systems belonging to a sender and a receiver of
information. This ambiguity (arguably) makes communication theory as we know it
in the context of such setups impossible, similarly to the way in which the
failure of the principle of mutually independent existence (arguably) makes
physical theory as we know it impossible.

Before beginning let me emphasise that the general approach represented by the
investigation into information causality is only one of a number of
principle-theoretic approaches that one can take regarding the question of how
to distinguish quantum from super-quantum theories. In the kind of approach
exemplified by the investigation into information causality, one focuses on
sets of static correlation tables associated with quantum and super-quantum
theories, and in particular one disregards the dynamics of (super-)quantum
systems. There is another family of principle-theoretic approaches to the
question, however, wherein a richer framework is considered that does include
dynamics.\footnote{For further references, as well as an accessible
  description of one of these reconstructions of quantum theory, see
  \citet[]{koberinski2018}.} \citeauthor[]{popescu1994}'s seminal
\citeyearpar[]{popescu1994} investigation is an example of the former type of
approach, though they themselves consider the latter, dynamical, approach to
have the potential for deeper insight. For my part I do not consider any
particular approach to be superior. Principle-theoretic approaches to the
characterisation of quantum theory augment our understanding of the world by
illuminating various aspects of it to us. Which particular aspect of the world
is illuminated by an investigation will depend upon the particular
question---and the framework which defines it---that is asked.\footnote{Thanks
  to Giulio Chiribella for expressing something like this statement in answer
  to a question posed to him at the workshop `Contextuality: Conceptual
  Issues, Operational Signatures, and Applications', held at the Perimeter
  Institute in July, 2017.} I am highly skeptical of the idea that any one
framework is sufficient by itself to illuminate all. Rather, these different
frameworks of analysis should be seen as conveying to us information---in
general neither literal nor complete---regarding different aspects of one and
the same reality.

The rest of this paper will proceed as follows: I will introduce
Popescu-Rohrlich (PR) correlations in \S\ref{sec:prcorr}. In \S\ref{sec:game} I
will introduce the `guessing game' by which the principle of information
causality is standardly operationally defined. The principle of information
causality itself will be introduced in \S\ref{sec:ic}, wherein I will also
describe how it can be used to derive the Tsirelson bound. I will argue in that
section that information causality has not been sufficiently motivated to play
the role of a foundational principle of nature, and in the remainder of the
paper I will consider how one might begin to provide it with such a
motivation. This analysis begins in \S\ref{sec:demopoulos} where I describe an
argument, due to Demopoulos, to the effect that the no-signalling condition can
be viewed as a generalisation, appropriate to an irreducibly statistical theory,
of Einstein's principle of mutually independent existence interpreted
as a constraint on physical practice. Then in \S\ref{sec:howposs} I argue that
a promising route toward successfully motivating information causality is to in
turn consider it as a further generalisation of no-signalling that is
appropriate to a theory of communication. I describe, however, some important
obstacles that must yet be overcome if the project of establishing information
causality as a foundational principle of nature is to succeed.

\section{Popescu-Rohrlich correlations}
\label{sec:prcorr}

Consider a correlated state $\sigma$ of two two-level
subsystems.\footnote{Elements of the exposition in this and the next section
  have been adapted from \citet[]{bub2012,bub2016} and \citet[]{pawlowski2009}.}
Let Alice and Bob each be given one of the subsystems, and instruct them to
travel to distinct distant locations. Let $p(A, B|a, b)$ be the probability that
Alice and Bob obtain outcomes $A$ and $B$, respectively, after measuring their
local subsystems with the respective settings $a$ and $b$. If $A,B \in \{\pm
1\}$, the expectation value of the outcome of their combined measurement is
given by: $$\langle a, b \rangle = \sum_{i, j \in \{1,-1\}}  (i \cdot j) \cdot
p(i, j|a, b),$$ where $A = i$ and $B = j$. Less concisely, this is:
\begin{align*}
\langle a, b \rangle & = 1 \cdot p(1,1|a, b) - 1 \cdot p(1,\text{-}1|a, b) - 1
\cdot p(\text{-}1,1|a, b) + 1 \cdot p(\text{-}1,\text{-}1|a, b) \\
& = p(\mbox{same}|a, b) - p(\mbox{different}|a, b).
\end{align*}
Since $p(\mbox{same}|a, b)$ + $p(\mbox{different}|a, b)$ = 1, it follows that
$\langle a, b \rangle$ + $2 \cdot p(\mbox{different}|a, b)$ = 1, so
that: $$p(\mbox{different}|a, b) = \frac{1 - \langle a, b \rangle}{2}.$$
Similarly, we have that $$p(\mbox{same}|a, b) = \frac{1 + \langle a, b
  \rangle}{2}.$$

Now imagine that $\sigma$ is such that the probabilities for the results of
experiments with settings $a, b, a', b'$, where $a'$ and $b'$ are different from
$a$ and $b$ but arbitrary \citep[p. 382]{popescu1994}, are:
\begin{align}
  \label{eqn:prprobs}
  p(1,1|a,b) & = p(\text{-}1,\text{-}1|a,b) = 1/2, \nonumber \\
  p(1,1|a,b') & = p(\text{-}1,\text{-}1|a,b') = 1/2, \nonumber \\
  p(1,1|a',b) & = p(\text{-}1,\text{-}1|a',b) = 1/2, \nonumber \\
  p(1,\text{-}1|a',b') & = p(\text{-}1,1|a',b') = 1/2.
\end{align}
In other words, if at least one of their settings is one of $a$ or $b$, then
Alice's and Bob's results are guaranteed to be the same. Otherwise they are
guaranteed to be different. These correlations are called `PR' correlations
after \citet{popescu1994}.

Alice's marginal probability $p(1_A|a,b)$ of obtaining the outcome 1 given
that she measures $a$ and Bob measures $b$ is defined as: $p(1_A,1_B|a,b)$ +
$p(1_A,\text{-}1_B|a,b)$. The no-signalling condition requires that her marginal
probability of obtaining 1 is the same irrespective of whether Bob measures $b$
or $b'$, i.e. that $p(1_A|a,b)$ = $p(1_A|a,b')$, in which case we can write her
marginal probability simply as $p(1_A|a)$. In general, no-signalling requires
that
\begin{align}
  \label{eqn:nosig}
  p(A|a,b) & = p(A|a,b'), & p(A|a',b) & = p(A|a',b'), \nonumber \\
  p(B|a,b) & = p(B|a',b), & p(B|a, b') & = p(B|a', b').
\end{align}
The reader can verify that the PR correlations \eqref{eqn:prprobs} satisfy
the no-signalling condition \eqref{eqn:nosig}.

If we imagine trying to simulate the PR correlations \eqref{eqn:prprobs} with
some bipartite general non-signalling system $\eta$, then the probability of a
successful simulation (assuming a uniform probability distribution over the
possible joint measurements $(a,b)$, $(a,b')$, $(a',b)$, and $(a',b')$) is given
by:\footnote{By a `successful simulation' I mean a single joint measurement in
  which Alice and Bob get opposite outcomes---(1,-1) or (-1,1)---if their
  settings are $(a', b')$, or the same outcome---(1,1) or (-1,-1)---otherwise.}
\begin{align*}
  \frac{1}{4}\big(p(\mbox{same}|a,b) + p(\mbox{same}|a,b') +
  p(\mbox{same}|a',b) + p(\mbox{different}|a',b')\big) \\
  = \frac{1}{4}\Bigg(\frac{1 + \langle a, b \rangle}{2} + \frac{1 + \langle
    a, b' \rangle}{2} + \frac{1 + \langle a', b \rangle}{2} + \frac{1 - \langle
    a', b' \rangle}{2} \Bigg) \\
  = \frac{1}{2}\Bigg(1 + \frac{\langle a, b \rangle + \langle a, b' \rangle +
    \langle a', b \rangle - \langle a', b' \rangle}{4}\Bigg).
\end{align*}
Notice that $\langle a, b \rangle + \langle a, b' \rangle + \langle a', b
\rangle - \langle a', b' \rangle$ is just the Clauser-Horne-Shimony-Holt (CHSH)
correlation expression \citep[]{chsh1969}. So the probability of a successful
simulation of the PR correlations by $\eta$ is:
\begin{align}
  \label{eqn:succsim}
  p(\mbox{successful sim}) = \frac{1}{2}\Bigg(1 + \frac{\mbox{CHSH}}{4}\Bigg),
\end{align}
with CHSH = 4 if $\eta$ is itself a PR-system.\footnote{The reader may be
  familiar with the use of the term `PR-box' to refer to systems whose
  subsystems are correlated as in \eqref{eqn:prprobs}. I find the term `box' to
  be misleading since it conveys the idea of a spatially contiguous region
  occupied by a combined system. Bub's \citeyearpar[]{bub2016} banana imagery is
  far less misleading in this sense. Below I will not use figurative language at
  all, but will (boringly) refer merely to such entities as `PR-systems',
  `PR-correlated systems', and so on.} As is well known, classically correlated
systems are bounded by $|\mbox{CHSH}| \leq 2$. Thus the optimum probability of
simulating PR correlations with a bipartite classical system is given by 1/2(1 +
2/4) = 3/4. Quantum correlations are bounded by $|\mbox{CHSH}| \leq 2\sqrt 2$.

\section{Alice and Bob play a guessing game}
\label{sec:game}

At this point it will be convenient to change our notation. From now on I will
refer to the measurement settings $a$ and $a'$ as 0 and 1, respectively, and
likewise for $b$ and $b'$. The outcomes 1 and -1 will also be respectively
relabelled as 0 and 1. This will allow us to describe PR correlations more
abstractly using the exclusive-or (alternately: modulo two addition) operator as
follows:
\begin{align}
  \label{eqn:xorpr}
  M_1 \oplus M_2 = m_1 \cdot m_2
\end{align}
where capital letters refer to measurement outcomes and small letters to
measurement settings. To illustrate, for a given 01-experiment (formerly
$(a,b')$) there are two possible outcomes: 00 and 11 (formerly: (1,1) and
(-1,-1)), and we have: $0 \oplus 0 = 0 \cdot 1$ and $1 \oplus 1 = 0 \cdot 1$,
respectively.

Now imagine the following game. At the start of each round of the game, Alice
and Bob receive random and independently generated bit strings $\mathbf{a} =
a_{N-1},a_{N-2},\dots,a_0$ and $\mathbf{b} = b_{n-1},b_{n-2},\dots,b_0$,
respectively, with $N = 2^n$. They win a round if Bob is able to guess the value
of the $\textbf{b}^{\mbox{\scriptsize th}}$ bit in Alice's list. For example,
suppose Alice receives the string $a_{7}a_{6}a_{5}a_{4}a_{3}a_{2}a_{1}a_{0}$,
and Bob receives the string 110. Then Bob must guess the value of $a_{6}$. They
win the game if Bob is able to guess correctly over any sequence of rounds.

Besides this the rules of the game are as follows. Before the game starts, Alice
and Bob are allowed to determine a mutual strategy and to prepare and share
non-signalling physical resources such as classically correlated systems, or
quantum systems in entangled states, or PR-systems, or other (bipartite) systems
manifesting non-signalling correlations. They then go off to distinct distant
locations, taking with them their portions of whatever systems were previously
prepared. Once separated, Alice receives her bit string $\mathbf{a}$ and Bob his
bit string $\mathbf{b}$. She is then allowed to send Bob one additional
classical bit $c$, upon receipt of which Bob must guess the value of Alice's
$\mathbf{b}^{\mbox{\scriptsize th}}$ bit.

Alice and Bob can be certain to win the game if they share a number of
PR-systems. I will illustrate the case of $N=4$, which requires three
PR-systems (per round) labelled \textbf{I}, \textbf{II}, and \textbf{III}. Upon
receiving the bit string $\mathbf{a} = a_3a_2a_1a_0$, Alice measures $a_0 \oplus
a_1$ on her part of system \textbf{I} and gets the result $A_I$. She then
measures $a_2 \oplus a_3$ on her part of system \textbf{II} and gets the outcome
$A_{II}$. She then measures $(a_o \oplus A_I) \oplus (a_2 \oplus A_{II})$ on her
part of system \textbf{III} and gets the result $A_{III}$. She finally sends $c
= a_0 \oplus A_I \oplus A_{III}$ to Bob. Meanwhile, Bob, who has previously
received $\mathbf{b} = b_1b_0$, measures $b_0$ on his parts of systems
\textbf{I} and \textbf{II}, and gets back the results $B_I$ and $B_{II}$. He
also measures $b_1$ on system \textbf{III} with the result $B_{III}$.

Bob's next step depends on the value of $\mathbf{b}$, i.e. on which of Alice's
bits he has to guess. When $\mathbf{b} = b_1b_0 = 00$ (i.e. when Bob must guess
the 0$^{\mbox{\scriptsize th}}$ bit) or $\mathbf{b} = b_1b_0 = 01$ (i.e. when
Bob must guess the 1$^{\mbox{\scriptsize st}}$ bit) his guess should be:
\begin{align}
  \label{eqn:guess0or1}
  c \oplus B_{III} \oplus B_I = a_0 \oplus A_I \oplus A_{III} \oplus B_{III}
  \oplus B_I.
\end{align}
For since $A_{III} \oplus B_{III} = \big((a_0 \oplus A_I) \oplus (a_2 \oplus
A_{II})\big) \cdot b_1$, we have:
\begin{align}
  \label{eqn:b1equal0}
  & a_0 \oplus A_I \oplus A_{III} \oplus B_{III} \oplus B_I \nonumber \\
  =\mbox{ } & a_0 \oplus A_I \oplus b_1(a_0 \oplus A_I) \oplus b_1(a_2 \oplus
  A_{II}) \oplus B_I \nonumber \\
  =\mbox{ } & a_0 \oplus A_I \oplus B_I \nonumber \\
  =\mbox{ } & a_0 \oplus b_0(a_0 \oplus a_1).
\end{align}
If $\mathbf{b} = 00$ then \eqref{eqn:b1equal0} correctly yields $a_0$. If
$\mathbf{b} = 01$ then \eqref{eqn:b1equal0} correctly yields $a_1$.

Suppose instead that $\mathbf{b} = 10$ or $\mathbf{b} = 11$. In this
case, Bob's guess should be
\begin{align}
  \label{eqn:guess2or3}
  c \oplus B_{III} \oplus B_{II} = a_0 \oplus A_I \oplus A_{III} \oplus
  B_{III} \oplus B_{II}.
\end{align}
This is
\begin{align}
  \label{eqn:b1equal1}
  =\mbox{ } & a_0 \oplus A_I \oplus b_1(a_0 \oplus A_I) \oplus b_1(a_2 \oplus A_{II})
  \oplus B_{II} \nonumber\\
  =\mbox{ } & (a_0 \oplus A_I) \oplus (a_0 \oplus A_I) \oplus (a_2 \oplus
  A_{II}) \oplus B_{II} \nonumber\\
  =\mbox{ } & a_2 \oplus A_{II} \oplus B_{II} \nonumber\\
  =\mbox{ } & a_2 \oplus b_0(a_2 \oplus a_3).
\end{align}
If $\mathbf{b} = 11$ then \eqref{eqn:b1equal1} correctly yields $a_3$. If
$\mathbf{b} = 10$ then \eqref{eqn:b1equal1} correctly yields $a_2$.

In general, given $N-1$ PR-correlated systems per round,\footnote{These are to
  be arranged in an inverted pyramid so that the results of Alice's
  (respectively, Bob's) local measurements on the first $2^{n-1}$ PR-systems
  are used to determine the local settings for her (his) next $2^{n-2}$
  measurements, and so on, for $(n-i) \geq 0$. Note that the cost in the number
  of PR-systems needed scales exponentially with respect to the length of
  $\mathbf{b}$. I will return to this point later.} and a single classical bit
per round communicated by Alice to Bob, Alice and Bob can be certain to win the
game for any value of $N$. In other words, given these resources and a single
classical bit communicated to him by Alice, Bob can access the value of any
single bit from her data set, however large that data set is. This result
further generalises to the case where Alice is allowed to send not just one but
$m$ bits $c_{m-1}\dots c_0$ to Bob in a given round, and Bob is required to
guess an arbitrary set of $m$ bits from Alice's data set. Note that if Alice is
not allowed to send anything to Bob, i.e., when $m$ = 0, then Bob will not be
able to access the values of any of Alice's bits irrespective of how many
PR-systems they share. This is a consequence of the fact that PR-correlations
satisfy the no-signalling principle \eqref{eqn:nosig}.

\section{Information causality and the Tsirelson bound}
\label{sec:ic}

As we saw in the last section, Alice and Bob can be certain to win the guessing
game described there if they share a number of PR-correlated systems prior to
going off to their respective locations. Note that if they do not use any
correlated resources, they can still be sure to win the occasional round if
Alice always sends Bob the value of whatever bit is at a previously agreed-upon
fixed position $a_k$ in her list. In this case, Bob will be guaranteed to guess
correctly whenever $\mathbf{b}$ singles out $k$ (but only then; otherwise he
must rely on blind luck). If Alice and Bob share a sequence of classically
correlated random bits, on the other hand, then Bob will be able to access the
value of a single in general different $a_i$ in Alice's list on each round.

Now consider the case where Alice and Bob share general no-signalling systems,
i.e. bipartite systems such that the correlations between their subsystems
satisfy the no-signalling condition. Recall that the probability that a
non-signalling system simulates a PR-system on a given run depends on the value
of CHSH in \eqref{eqn:succsim} that is associated with it. For convenience we
will define $E =_{\mathit{df}} \mbox{CHSH}/4$ so that \eqref{eqn:succsim}
becomes:
\begin{align}
  \label{eqn:succsim2}
  p(\mbox{successful sim}) = \frac{1}{2}(1 + E).
\end{align}
When $E = 1$ for a given non-signalling system, then it just is a PR-system, and
the probability of a successful simulation is 1. When $E < 1$, then for given
settings $m_1, m_2$, the values of the outcomes $M_1, M_2$, will in general not
satisfy the relation \eqref{eqn:xorpr}, i.e. $M_1 \oplus M_2$ will not always
equal $m_1 \cdot m_2$. For a given attempted simulation, let us say that $M_2$
is `correct' whenever \eqref{eqn:xorpr} holds, and `incorrect'
otherwise.\footnote{There is of course no reason why we should not say that
  $M_1$ rather than $M_2$ is incorrect, but for the analysis that follows it
  is convenient to take Bob's point of view.}

Recall that in the $N=4$ game above, at the end of each round, Bob guesses
either (i) $c \oplus B_{III} \oplus B_{I}$, or (ii) $c \oplus B_{III} \oplus
B_{II}$, depending on the value of $\mathbf{b}$. We will consider only case (i),
as the analysis is similar for (ii). If both $B_I$ and $B_{III}$ are `correct',
then for that particular round, the non-signalling systems will have yielded the
same guess for Bob as PR-systems would have yielded:
\begin{align}
  \label{eqn:prmatch}
  (c \oplus B_{III} \oplus B_{I})_{NS} = (c \oplus B_{III} \oplus B_{I})_{PR}.
\end{align}
Note that if \emph{both} $B_I$ and $B_{III}$ are \emph{incorrect},
\eqref{eqn:prmatch} will still hold, since in general $x_1 \oplus x_2 =
\overline{x_1} \oplus \overline{x_2}$. So either way Bob will guess right. The
probability of an unsuccessful simulation is $$1-\frac{1}{2}(1+E) =
\frac{1}{2}(1-E).$$ Thus the probability that Bob makes the right guess on a
given round in the $N=4$ game is:
\begin{align*}
\left(\frac{1}{2}(1 + E)\right)^2 + \left(\frac{1}{2}(1 - E)\right)^2 =
\frac{1}{2}(1+E^2).
\end{align*}
In the general case, for $N = 2^n$, one can show
\citep[]{pawlowski2009,bub2012,bub2016} that the probability that Bob correctly
guesses Alice's $\mathbf{b}^{\mbox{\scriptsize th}}$ bit is
\begin{align}
  \label{eqn:prguess}
  p_{\mathbf{b}} = \frac{1}{2}(1 + E^n).
\end{align}

The binary entropy $h(p_{\mathbf{b}})$ associated with $p_{\mathbf{b}}$ is given
by $$h(p_{\mathbf{b}}) = \text{-}p_{\mathbf{b}}\log_2{p_{\mathbf{b}}} - (1 -
p_{\mathbf{b}})\log_2{(1 - p_{\mathbf{b}})}.$$ In the case where Bob has no
information about Alice's $\mathbf{b}^{\mbox{\scriptsize th}}$ bit,
$p_{\mathbf{b}} = 1/2$ and $h(p_{\mathbf{b}}) = 1$. If Alice then sends Bob $m$
bits, then in general Bob's information about that bit will increase by some
non-zero amount. \citet[]{pawlowski2009} propose the following constraint on
this quantity, which they call the `information causality' principle:

\begin{quote}
The information gain that Bob can reach about a previously unknown to him data
set of Alice, by using all his local resources and $m$ classical bits
communicated by Alice, is at most $m$ bits
\citeyearpar[p. 1101]{pawlowski2009}.
\label{quo:ic}
\end{quote}

For example, assuming that the $N = 2^n$ bits in Alice's bit string $\mathbf{a}$
are unbiased and independently distributed, then if Alice sends Bob a single bit
(i.e. when $m = 1$), information causality asserts that Bob's information about
the $\mathbf{b}^{\mbox{\scriptsize th}}$ bit in Alice's string may increase by
no more than $1/2^n$, i.e.,
\begin{align}
  \label{eqn:infcaus}
  h(p_{\mathbf{b}}) \geq 1 - \frac{1}{2^n}.
\end{align}
As \citet[]{pawlowski2009} show, the principle is satisfied within quantum
mechanics. But within any theory which permits correlations with a value of $E$
exceeding $1/\sqrt{2}$ (i.e. any theory which allows correlations above the
Tsirelson bound), one can find an $n$ such that for a given $m$ the principle
is violated (for example, let $E = .72$, $m = 1$, and $n = 10$).\footnote{Note
  that when $E = 1$ the principle is always violated for any $m$ and
  $n$.}$^{\mbox{,}}$\footnote{I have followed Bub in expressing information
  causality as a constraint on binary entropy, as conceptually this is a more
  transparent way of expressing \citeauthor[]{pawlowski2009}'s `qualitative'
  statement of the principle in terms of concrete information-theoretic
  quantities. While \citeauthor[]{pawlowski2009} also relate information
  causality to the binary entropy \citeyearpar[p. 1102 and Supplementary
  Information \S{III}]{pawlowski2009}, their general results (that
  information causality is satisfied within quantum mechanics and that it is
  violated within any theory which allows correlations above the Tsirelson
  bound) begin with the formulation of information causality as a condition
  on mutual information rather than binary entropy. For our purposes it is
  immaterial which formulation one chooses; in particular,
  \citet[\S\S{11.4--11.5}]{bub2012} has shown that \eqref{eqn:infcaus} is
  entailed by \citeauthor[]{pawlowski2009}'s formulation and moreover proves
  that \eqref{eqn:infcaus} is satisfied when $E = \frac{1}{\sqrt 2}$.}

Given that any correlations above the Tsirelson bound will demonstrably violate
the principle in this sense, it is tempting to view information causality as
the answer to the question (i) of why nature does not allow correlations above
this bound. And since the Tsirelson bound represents the maximum value of the
CHSH expression for quantum correlations, one is further tempted to view
information causality as the answer to the question (ii) of why only quantum
correlations are allowable in nature. Indeed, \citeauthor[]{pawlowski2009}
suggest that information causality ``might be one of the foundational
properties of nature'' \citeyearpar[p. 1101]{pawlowski2009}.

There is a subtlety here, however. The set of quantum correlations forms a
convex set which can be represented as a multi-dimensional region of points such
that the points within this region that are furthest from the centre are at the
Tsirelson bound \citep[\S{}5.1]{bub2016}. Information causality disallows
correlations beyond this bound, as we saw. It also disallows some correlations
below the bound that are outside of the quantum convex set \citep[for a
  discussion, see][]{pawlowski2016}. However there is numerical evidence that
there exist correlations within the bound but outside of the quantum convex set
that satisfy the information causality principle \citep[]{navascues2015}. So it
appears unlikely (though this was not known in 2009) that information causality
can provide an answer to question (ii). It nevertheless remains promising as a
principle with which to answer question (i) and can arguably still be thought of
as a fundamental principle in that sense. Analogously, the fact that
super-quantum no-signalling correlations are possible does not, in itself,
undermine the status of no-signalling as a fundamental principle.

The information causality principle must be given some independent motivation if
it is to play this explanatory role, however. For even a conventionalist would
agree that some stipulations are better than others \citep[]{disalle2002}. Thus
some independent reason should be given for why one might be inclined to accept
the principle. Of course, the statement that the communication of $m$ bits can
yield no more than $m$ bits of additional information to a receiver about a data
set unknown to him is an intuitive one. But foundational principles of nature
should require more for their motivation than such bare appeals to
intuition. After all, quantum mechanics, which the principle aims to legitimate,
arguably already violates many of our most basic
intuitions. \citet[]{pawlowski2009} unfortunately do not say very much to
motivate information causality. But two ideas can be gleaned from statements
made in their paper. The first is that in a world in which violations of
information causality could occur, ``certain tasks [would be] `too simple'''
(p. 1101). The second is that in such a world there would be ``implausible
accessibility of remote data'' (ibid.). The former idea has been expressed in
this general context before. Van Dam \citeyearpar[]{vanDam2013}, notably, shows
that in a world in which PR-correlations exist and can be taken advantage of,
only a trivial amount of communication (i.e. a single bit) is required to
perform any distributed computational task. Van Dam argues (ibid., p. 12) that
this is a reason to believe that such correlations cannot exist, for they
violate the principle that ``Nature does not allow a computational `free
lunch''' (ibid., p. 9).\footnote{Cf. \citet[][]{aaronson2005a}.}
\citet[pp. 180-181]{bub2012} echoes this thought by listing examples of
distributed tasks (`the dating game' and `one-out-of-two' oblivious transfer)
which would become implausibly trivial if PR-correlated systems could be used.

Later in this paper I will argue that although such statements are
unsatisfactorily vague, they nevertheless get at something that is importantly
right; although they are right in, perhaps, a different sense than their authors
envision. For now let me just say that even if one accepts van Dam's argument
that pervasive trivial communication complexity is implausible and should be
ruled out---and that this should constitute a constraint on physical
theory---not all correlations above the Tsirelson bound in fact result in the
trivialisation of communication complexity theory.\footnote{Communication
  complexity theory aims to quantify the communicational resources---measured
  in transmitted bits---required to solve various distributed computational
  problems. A good reference work is that of \citet[]{hushilevitz1997}.}
\citet[]{brassard2006} have extended van Dam's result by showing that
(probabilistic) pervasive trivial communication complexity can be achieved for
values of $E > \sqrt{6}/3$. But this still leaves a range of values for $E$
open; physical correlations with associated values of $E$ between the quantum
mechanical maximum of $1/\sqrt 2$ and $\sqrt{6}/3$ have not been shown to result
in pervasive trivial communication complexity and cannot---at least not yet---be
ruled out on those grounds. Thus the avoidance of pervasive trivial
communication complexity cannot be used to motivate information causality in the
way suggested by the statements of \citet[]{pawlowski2009}. In fairness, to say
as they do that certain tasks would be `too simple' in a world in which
information causality is violated is not the same as saying that they would be
trivial. The task remains, then, of expressing more precisely what is meant by
`too simple' in a way that is sufficient to motivate ruling out theories which
violate the information causality principle in a less than maximal way (in
particular with a value of $E \leq \sqrt{6}/3$). We will return to this point
later.

Regarding their second idea---that a world in which information causality is
violated would manifest ``implausible accessibility of remote data''
(p. 1101)---\citet[]{pawlowski2009} again do not say
enough,\footnote{\citet[p. 429]{pawlowski2016} do expand on the idea of
  implausible accessibility slightly: ``we have transmitted only a single bit
  and the PR-boxes are supposed to be no-signalling so they cannot be used to
  transmit the other. Somehow the amount of information that the lab of Bob
  has is larger than the amount it received. Things like this should not
  happen.'' I do not think this adds anything substantial to the idea
  expressed by \citet[]{pawlowski2009} that such a situation is
  `implausible'.} although the idea is perhaps alluded to implicitly in another
assertion they (too briefly) make, namely that information causality
generalises the no-signalling principle (ibid., p. 1103). We will come back to
this point later. In any case, the idea of implausible accessibility is
fortunately expanded upon by \citet[]{bub2012}, who motivates it in the
following way:

\begin{quote}
when the bits of Alice's data set are unbiased and independently distributed,
the intuition is that if the correlations can be exploited to distribute one bit
of communicated information among the $N$ unknown bits in Alice's data set, the
amount of information distributed should be no more than $\frac{1}{N}$ bits,
because there can be no information about the bits in Alice's data set in the
previously established correlations themselves (p. 180).
\end{quote}

Partly for this reason, Bub argues that the principle is misnamed. Drawing on
the idea of implausible accessibility he argues that `information causality'
should rather be referred to as information \emph{neutrality}: ``The principle
really has nothing to do with causality and is better understood as a
\emph{constraint on the ability of correlations to enhance the information
  content of communication in a distributed task}'' (ibid., emphasis in
original). Bub reformulates the principle as follows:

\begin{quote}
Correlations are informationally neutral: insofar as they can be exploited to
allow Bob to distribute information communicated by Alice among the bits in an
unknown data set held by Alice in such a way as to increase Bob's ability to
correctly guess an arbitrary bit in the data set, they cannot increase Bob's
information about the data set by more than the number of bits communicated by
Alice to Bob (ibid.).
\end{quote}

Stated in this way the principle sounds plausible and seems, intuitively, to be
correct. However if the principle is to be of aid in ruling out classes of
physical theory then it should be more than just intuitively plausible. If the
goal of answering the question `Why the Tsirelson bound?' is to give a
convincing reason why correlations that are above the bound should be regarded
as impossible, then if the fact that such correlations violate informational
neutrality is to be one's answer, one should give an independent motivation for
why correlations must be informationally neutral. One might, for instance,
motivate information neutrality by showing how it generalises or gives
expression in some sense to a deeper underlying principle that is already
well-motivated, or by pointing to `undesirable consequences' of its failure. The
consequence of a `free computational lunch' given the existence of correlations
above the bound, if it could be demonstrated, could (perhaps) constitute an
example of the latter kind of motivation.

This said, there is a different way to think of the question `Why the Tsirelson
bound?' for which Bub's explication of information causality in terms of
informational neutrality is both a full answer and indeed an illuminating and
useful one. In this sense the question represents a desire to understand what
the Tsirelson bound expresses about correlations which violate it. Information
neutrality answers this question by directing attention to a feature that no
correlations above the bound can have. This feature, moreover, is one that we
can easily grasp and explicitly connect operationally with our experience of
correlated physical systems. On such a reading of the question, to answer
`information neutrality' is not of course to rule out that the world could
contain non-informationally-neutral physical correlations. But on this view
ruling out such a possibility is not the point, which is rather to provide a
physically meaningful principle to help us to understand what our current
physical theories, assuming they are to be believed, are telling us about the
structure of the world.

In the remainder of this paper, however, I will continue to consider the
information causality/neutrality principle as a possible answer in the first
sense to the question `Why the Tsirelson bound?'. I will continue to consider,
that is, whether there is some independent way of motivating the conclusion that
correlations which violate the condition should be ruled out.

\section{The `being-thus' of spatially distant things}
\label{sec:demopoulos}

Our goal is to determine whether there is some sense in which we can motivate
the idea that information causality must be satisfied by all physical theories
which treat of correlated systems. I will now argue that some insight into this
question can be gained if we consider the analogous question regarding
no-signalling. As I mentioned earlier, the no-signalling condition
\eqref{eqn:nosig} is not a relativistic constraint per se---in itself it is
merely a restriction on the marginal probabilities associated with experiments
on the subsystems of combined systems---but its violation entails the ability to
instantaneously signal, which is in tension if not in outright violation of the
constraints imposed by relativistic theory.\footnote{For a discussion of
  signalling in the context of special and general relativity see
  \citet[Ch. 4]{maudlin2011}.} Indeed, the independently confirmed relativity
theory can in this sense be thought of as an external motivation for thinking of
the no-signalling principle as a constraint on the marginal probabilities
allowable in any physical theory.

There is an arguably deeper way to motivate no-signalling, however, that can be
drawn from the work of Einstein and which has been expanded upon by
Demopoulos.\footnote{This is done in Demopoulos's monograph \emph{On
    Theories}; see fn. \ref{fn:demo}.} In the course of expressing his
dissatisfaction with the `orthodox' interpretation of quantum theory, Einstein
described two foundational ideas---what Demopoulos calls \emph{local realism}
and \emph{local action}. Realism in general, for Einstein, is a basic
presupposition of any physical theory. It amounts to the claim that things in
the world exist independently of our capability of knowing them; i.e.

\begin{quote}
the concepts of physics refer to a real external world, i.e., ideas are posited
of things that claim a `real existence' independent of the perceiving subject
(bodies, fields, etc.), and these ideas are, on the other hand, brought into as
secure a relationship as possible with sense impressions
(\citealt[]{einstein1948}, as translated by \citealt[p. 187]{howard1985}).
\end{quote}

\emph{Local} realism---alternately: the `mutually independent existence' of
spatially distant things---is the idea that things claim independent existence
from one another insofar as at a given time they are located in different parts
of space. Regarding this idea, Einstein writes:

\begin{quote}
Without such an assumption of the mutually independent existence (the `being
thus') of spatially distant things, an assumption which originates in everyday
thought, physical thought in the sense familiar to us would not be possible
(ibid.).
\end{quote}

In the concrete context of a physical system made up of two correlated subsystems
$S_1$ and $S_2$ (such as that described in the thought experiment of
\citealt[]{epr1935}), local realism requires that

\begin{quote}
every statement regarding $S_2$ which we are able to make on the basis of a
complete measurement on $S_1$ must also hold for the system $S_2$ if, after all,
no measurement whatsoever ensued on $S_1$ (\citealt[]{einstein1948}, as
translated by \citealt[p. 187]{howard1985}).
\end{quote}

In other words the value of a measurable theoretical parameter of $S_2$ must
not depend on whether a measurement is made on a system $S_1$ that is located
in some distant region of space. (And of course it must also not depend upon
the \emph{kind} of measurement performed on $S_1$;
cf. \citealt[][p. 186]{howard1985}.) Demopoulos notes that local realism as it
is applied in such a context is a condition imposed on the measurable
properties of the theory and hence it is a condition that is imposed at a
theory's `surface' or operational level. This is an important point that I will
return to later.

In the same \emph{Dialectica} article Einstein also formulated a second
principle:

\begin{quote}
For the relative independence of spatially distant things (A and B), this idea
is characteristic: an external influence on A has no \emph{immediate} effect on
B; this is known as the `principle of local action' ... The complete suspension
of this basic principle would make impossible the idea of the existence of
(quasi-) closed systems and, thereby, the establishment of empirically testable
laws in the sense familiar to us (\citealt[]{einstein1948}, as translated by
\citealt[p. 188]{howard1985}).
\end{quote}

The thought expressed in the second part of this statement seems similar to
Einstein's earlier assertion that `physical thought' would not be possible
without the assumption of local realism. However Demopoulos convincingly argues
that the principle of local realism, though it receives support from the
principle of local action, is a conceptually more fundamental principle than
the latter. For conceivably the principle of local realism---i.e. of `mutually
independent existence'---could be treated as holding, Demopoulos argues, even
in the absence of local action. Indeed this is so in Newtonian mechanics. For
example, Corollary VI to the laws of motion \citep[p. 423]{newton1999} states
that a system of bodies moving in any way whatsoever with respect to one
another will continue to do so in the presence of equal accelerative forces
acting on the system along parallel lines. This makes it possible to treat the
system of Jupiter and its moons, for example, as a quasi-closed system with
respect to the sun. For owing to the sun's great distance (and relative size),
the actions of the forces exerted by it upon the Jovian system will be
approximately equal and parallel. Corollary VI, moreover, is used by Newton to
prove Proposition 3 of Book I \citep[p. 448]{newton1999}, which enables one to
distinguish forces that are internal to a given system from forces that are
external to it, and which provides a criterion (i.e. that the motions of the
bodies comprising a system obey the Area Law with respect to its centre of
mass) for determining when the gravitational forces internal to a system have
been fully characterised. Thus despite its violation of local action,
Demopoulos argues convincingly that Einstein would not (or anyway should not)
have regarded a theory such as Newtonian mechanics as unphysical. It is still a
basic \emph{methodological} presupposition of Newtonian mechanics that
spatially distant systems have their own individual `being thus-ness', the
description of which is made possible via the theory's characteristic
methodological tool of successive approximation, in turn made possible by, for
example, Corollary VI, Proposition 3, and the notion of quasi-closed system
implied by them.\footnote{Demopoulos does not specifically mention either
  Corollary VI or Proposition 3 in his discussion, but I take them to be
  implicit therein. For a detailed analysis of Newton's method of successive
  approximations and the methodological role therein played by Corollary VI
  and Proposition 3, see \citet[]{harper2011}. For a discussion of the same
  in relation to general relativity, see \citet[]{disalle2006, disalle2016}.}

Einstein's principle of local realism or mutually independent existence
presupposes the framework of classical physics, which itself presupposes the
framework of classical probability theory. Demopoulos argues, however, that the
conceptual novelty of quantum theory consists in the fact that it is an
`irreducibly statistical theory', precisely in the sense that its probability
assignments, unlike those described by classical probability theory, cannot in
general be represented as weighted averages of two-valued measures over the
Boolean algebra of all possible properties of a physical system \citep[see
  also][]{pitowsky1989, pitowsky2006, dickson2011}. This raises the question of
whether one can formulate a generalisation of the mutually independent existence
condition that is appropriate for an irreducibly statistical theory such as
quantum mechanics.\footnote{I am not claiming here that Einstein himself would
  have been inclined to follow this line of reasoning.}

Recall that Einstein's mutually independent existence condition is a condition
that is imposed on the level of the measurable parameters of a theory and hence
at its `surface' or operational level. It requires, in particular, that the
value of a measurable property of a system $S_1$ in some region of physical
space $R_1$ is independent of what kind of measurement (or whether any
measurement) is performed on some system $S_2$ in a distant region of space
$R_2$, irrespective of whether $S_1$ and $S_2$ have previously interacted.

Demopoulos argues that in the context of an irreducibly statistical theory such
as quantum mechanics, it is in fact the no-signalling condition which
generalises the mutually independent existence condition. It does so in the
sense that like mutually independent existence, no-signalling is a
surface-level constraint on the local facts associated with a particular
system, requiring that these facts be independent of the local surface-level
facts associated with other spatially distant systems. Unlike the mutually
independent existence condition, however, these local facts refer to the
marginal probabilities associated with a system's measurable properties rather
than with what one might regard as those properties themselves. Specifically,
no-signalling asserts that the marginal probability associated with a
measurement on a system $S_1$ at a given location $R_1$ is independent of what
kind of measurement (or whether any measurement) is performed on some system
$S_2$ in a distant region of space $R_2$.\footnote{It is worth noting that the
  parameter independence condition (\citealt[]{shimony1993}) is just the
  no-signalling condition extended to include a hypothetical, possibly
  hidden, set of \emph{underlying} parameters.} In this way no-signalling
allows us to coherently treat systems in different regions of physical space as
if they had mutually independent existences---i.e. as quasi-closed systems in
the sense described above---and thus allows for the possibility of `physical
thought' in a methodological sense and for ``the establishment of empirically
testable laws in the sense familiar to us'' (\citealt[]{einstein1948}, as
translated by \citealt[p. 188]{howard1985}). Demopoulos argues that quantum
mechanics, even under its orthodox interpretation, is in this way legitimated
by the principle and may be thought of as a local theory of nonlocal
correlations.

\section{Mutually independent existence and communication}
\label{sec:howposs}

In the previous section we saw that no-signalling can be regarded as
generalising a criterion for the possibility of `physical thought' originally
put forward by Einstein. And we saw that since quantum mechanics satisfies
no-signalling, one may think of that theory, even under its orthodox
interpretation, as in this sense legitimated methodologically by the
principle. As we saw in \S\S\ref{sec:prcorr}-\ref{sec:ic}, however, other
conceivable physical theories---some of which allow for stronger-than-quantum
correlations---satisfy the no-signalling condition as well. In light of this,
`information causality' (or `information neutrality', in Bub's terminology) was
put forward by \citet[]{pawlowski2009} as an additional foundational principle
for more narrowly circumscribing the class of physically sensible theories. But
in \S\ref{sec:ic} I argued that the principle requires further motivation
before it can legitimately be seen as playing this role. With our recent
discussion of no-signalling in mind, let us now consider the proposal of
\citeauthor[]{pawlowski2009} again.

\emph{No-signalling} asserts that the marginal probabilities associated with
Alice's local measurements on a system $S_A$ in a region $R_A$ are independent
of what kind of measurement (or whether any measurement) is performed by Bob
locally on a system $S_B$ in a distant region $R_B$. \emph{Information
  causality} asserts that Bob can gain no more than $m$ bits of information
about Alice's data set if she sends him only $m$
bits. \citet[p. 1101]{pawlowski2009} remark that ``The standard no-signalling
condition is just information causality for $m = 0$''. \citet[p. 180]{bub2012}
considers this remark to be misleading, but presumably all that
\citeauthor[]{pawlowski2009} intend is that if Alice and Bob share
\emph{signalling} correlations, then Alice may provide Bob with information
about her data set merely by measuring it, i.e. without actually sending him
any bits. The information causality principle disallows this for any value of
$E$, as does no-signalling.\footnote{That is, for any value of $E$ within the
  allowed range of: $\text{-}1 \leq E \leq 1$.}

\begin{figure}[t]
\footnotesize
$$
\begin{array}{l | l | l || l | l}
  b_0 & a_2 & a_3 & a_2 \oplus a_3 & G \\ \hline
  0   & 0  & 0   & 0              & 0 \\ \hline
  0   & 0  & 1   & 1              & 0 \\ \hline
  0   & 1  & 0   & 1              & 1 \\ \hline
  0   & 1  & 1   & 0              & 1 \\ \hline
  1   & 0  & 0   & 0              & 0 \\ \hline
  1   & 0  & 1   & 1              & 1 \\ \hline
  1   & 1  & 0   & 1              & 0 \\ \hline
  1   & 1  & 1   & 0              & 1
\end{array}
$$
\caption{A summary of the possible outcomes associated with Bob's measurement
  $G$ (his `guess') in the guessing game of \S\ref{sec:game}, based on
  Eq. \eqref{eqn:b1equal1}. If all atomic variables are assumed to be equally
  likely to take on a value of 0 or 1, then $G$ is probabilistically independent
  of Alice's measurement setting $a_2 \oplus a_3$, but not of its components
  $a_2$ and $a_3$, since, for example, $p(G=0|a_2=0) = 3/4 \neq p(G=0|a_2=1)$,
  and $p(G=0|a_3=0) = 3/4 \neq p(G=0|a_3=1)$.}
\label{fig:wittprob}
\end{figure}

On the other hand when (for instance) $m = 1$, then in the case where they have
previously shared PR-correlated systems (i.e. systems such that $E = 1$), one
might argue that there arises a subtle sense in which the probabilities of Bob's
measurement outcomes can be influenced by Alice's remote measurement
settings. Consider the outcome of Bob's combined measurement $G =_{\mathit{df}}
c \oplus B_{III} \oplus B_{II}$, i.e. his `guess' \eqref{eqn:guess2or3}. From
\eqref{eqn:b1equal1} it would appear that Bob's outcome is in part determined by
the setting of Alice's measurement on system \textbf{II}, $a_2 \oplus a_3$,
since this appears explicitly in the equation. However in this case appearances
are misleading, for the reader can verify that $G$ is probabilistically
independent of $a_2 \oplus a_3$ (see figure \ref{fig:wittprob}). $G$ is
nevertheless probabilistically dependent on both of $a_2$ and $a_3$ considered
individually. So one might say that although the outcome of $G$ is not
influenced by any of Alice's measurement settings \emph{per se}, it does seem to
be influenced by the particular way in which those settings have been determined
(despite the fact that neither $a_2$ nor $a_3$ are directly used by Alice to
determine the value of the bit that she sends to Bob, $c$). Put a different way,
the constituents of Alice's measurement setting on system \textbf{II}
respectively determine the two possible outcomes of Bob's guess whenever he
performs the measurement $G$ (for a given $b_0$). Likewise in the case where Bob
measures $G' = c \oplus B_{III} \oplus B_I$ (i.e. his guess
\eqref{eqn:guess0or1}); the two possible outcomes of $G'$ are, respectively,
determined by the constituents of Alice's measurement settings on system
\textbf{I}, $a_0$ and $a_1$ (for a given $b_0$).

Note that since $a_2$ and $a_3$ (respectively: $a_0$ and $a_1$), besides being
the constituents of Alice's measurement settings on \textbf{II} (respectively:
\textbf{I}), are also in fact the values of bits in Alice's list $\mathbf{a}$,
the above considerations resonate with Bub's remark (quoted above) that
Tsirelson-bound-violating correlations are such that they may themselves include
information about Alice's data set in the context of a game like that described
in \S\ref{sec:game}. These considerations further suggest a sense, \emph{pace}
Bub, in which it could be argued that the name `information causality' is indeed
apt. For the bit of information $c$ that Alice sends to Bob can be thought of as
the `enabler' or `cause', at least in a metaphorical sense, of Bob's ability to
use this aspect of the correlations to his advantage
\citep[cf.][\S{}3.4]{pawlowski2016}.\footnote{Perhaps, though, a better name
  would be the `\emph{no} information causality' principle.}

Thus one can think of information causality as generalising no-signalling (in
the context of the protocol under which information causality is operationally
defined) in two ways. On the one hand information causality generalises
no-signalling in the sense alluded to by \citeauthor[]{pawlowski2009}; i.e. it
reduces to no-signalling for $m = 0$. On the other hand information causality
generalises no-signalling in the sense that, like the no-signalling principle,
it expresses a restriction on the accessibility of the remote measurement
settings of a distant party; but this restriction now applies not just to those
remote measurement settings themselves, but also more generally to the
components by which those measurement settings are determined. Since, as we saw
in the previous section, no-signalling is already well-motivated in the sense
that it gives expression within quantum mechanics to an arguably fundamental
assumption that is implicit in physical practice, the very fact that
information causality generalises no-signalling can be taken as a compelling
motivation for it.

Such a conclusion would be too quick, however, for it does not follow from the
fact that information causality generalises no-signalling that it continues to
give expression to the condition of mutually independent existence. But it is
mutually independent existence which, as we saw, motivates no-signalling as a
constraint on physical theories. Thus we must still ask whether a violation of
information causality would result in a violation of the mutually independent
existence condition in some relevant sense. Arguably this is indeed the
situation one is confronted with in the context of the guessing game described
above when it is played with Tsirelson-bound-violating correlated systems. On
the one hand, when Alice and Bob share maximally super-quantum systems
(i.e. PR-systems, for which $E = 1$), then after receiving $c$ there is a sense
in which Alice's system can be said to be `a part' of Bob's system in the
context of the game being played. For after receiving $c$ Bob has
\emph{immediate} access to the value of any single bit of Alice's that he would
like. Alice's bits may as well be his own for the purposes of the game. Indeed,
from this point of view the fact that the communication complexity associated
with any distributed computational task is trivial when PR-correlations are
used seems natural; for once Alice's and Bob's systems are nonlocally joined in
this way there is naturally no need for further communication. On the other
hand, when Tsirelson-bound-violating correlations that are non-maximal are
used, trivial communication complexity has not been shown to result in all
cases. But mutually independent existence is nevertheless violated in the sense
that the correlations shared prior to the beginning of the game, upon being
`activated' by Alice's classical message $c$ to Bob, contribute information
over and above $c$ to the information Bob then gains about Alice's data set;
they `implausibly' enhance the accessibility of Alice's data set by nonlocally
joining Alice to Bob, at least to some extent, in the sense just described.

Now it is one thing to claim that information causality gives expression to a
generalised sense of mutually independent existence. It is another, however, to
claim that mutually independent existence should be thought of as necessary in
this context. Recall that in the last section we saw that mutually independent
existence (arguably) must be presupposed if `physical thought' is to be
possible---in other words that it is (arguably) a fundamental presupposition
implicit in physical practice as such. And we saw that a form of this principle
holds in the context of Newtonian mechanics, which may be thought of as in that
sense a local theory of nonlocal forces. We also saw that a form of
mutually independent existence appropriate for an irreducibly statistical
theory---i.e. the no-signalling principle---holds in the context of quantum
mechanics, and that it may thus be thought of analogously as a local theory of
nonlocal correlations. The context of our current investigation is one which
involves considering communicating agents capable of building and manipulating
physical systems---thought of now as resources---for their own particular
purposes. Our context, that is, is the `practical' one associated with quantum
computation and information theory, recently described by \citet[]{cuffaro2017,
  cuffaroForthB}.\footnote{Similar ideas have been expressed by
  \citet{pitowsky1990, pitowsky1996, pitowsky2002}.} As Cuffaro has argued,
this context of investigation is in fact distinct from the more familiar
`theoretical' context that is associated with traditional foundational
investigations of quantum mechanics. A different way of putting this is that
quantum computation and information theory are `resource' or `control' theories
similarly to the science of thermodynamics \citep[]{myrvold2011, wallace2014,
  ladyman2018}. Thus the question of whether mutually independent existence is
necessary for the practice of quantum information and communication complexity
theory is a distinct question from the question of whether it is necessary for
physical practice in the traditional sense.

Without the presupposition of mutually independent existence---according to
which systems that occupy distinct regions of space are to be regarded as
existing independently of one another---the idea of a (quasi-) closed system
that can be subjected to empirical test, and in this sense `physical thought',
would not be possible (or anyway so argued Einstein). Analogously, one could
argue that in the context of a theory of communication---i.e. of the various
resource costs associated with different communicational protocols and their
interrelations---that it is necessary to presuppose that an operational
distinction can be made between the parties involved in a communicational
protocol. One might argue, that is, that it is constitutive of the very idea of
communication that it is an activity that takes place between what can be
effectively regarded as two mutually independently existing entities, and
moreover that such a distinction is presupposed when one quantifies the
complexity of a particular
protocol.\footnote{Cf. \citet[p. x]{hushilevitz1997}. Cf. also
  \citeauthor[]{maroney2018}'s \citeyearpar[]{maroney2018} emphasis on the
  initialisation and readout stages of an information processing task.} For
without the ability to make such an effective distinction between the systems
belonging to the sender and the receiver of information, it is not at all
obvious how one should begin to quantify the amount of information that is
required to be sent \emph{from} Alice \emph{to} Bob in the context of a
particular protocol. From this point of view it is indeed not surprising that
communication complexity theory becomes impossible (in the sense that all
communicational problems become trivially solvable) when PR-correlated systems
are available to use.

\section{Objections}
\label{sec:obj}

An objection to this line of thought is the following. Cannot something similar
be said in the context of the information causality game when Alice and Bob
share an entangled quantum system? For arguably \citep[cf.][]{howard1989} Alice
and Bob will become likewise inseparable or `nonlocally joined' in such a
scenario. And yet no one imagines the very possibility of the sciences of
quantum information theory and quantum communication complexity to have been
undermined as a result. So why should one believe them to be undermined by the
possibility of sharing systems whose correlations violate the Tsirelson bound?
This objection, however, involves a description of the situation regarding the
sharing of an entangled quantum system that is below the surface-level
characterisation that is relevant to our discussion. It therefore does not
undermine the considerations of the previous section.

Consider the description of a classical bipartite communication protocol. Both
before and after communication has taken place, such a description may be
regarded as decomposable into three parts: a sending system, a receiving
system, and something communicated between them. For a quantum protocol the
possibility of such a decomposition is in general far less obvious as a result
of the well-known conceptual intricacies associated with entangled quantum
states. However whether or not Alice and her system, and Bob and his system,
are `in reality' inseparably entangled with one another, it remains the case,
both before (because of quantum mechanics' satisfaction of the no-signalling
condition) and after the communication of a classical message (because of
quantum mechanics' satisfaction of the information causality condition), that
Alice's system, Bob's system, and the message $c$ may be operationally
distinguished from one another in the sense that Bob cannot take advantage of
the underlying connection he has with Alice and her system via the correlations
he shares with her to gain information about her data set over and above what
has been provided to him via $c$. It is true that previously shared quantum
correlations enable one to communicate with greater efficiency than is possible
using only previously shared classical correlations. As \eqref{eqn:prguess}
shows, Bob has a higher probability of guessing correctly in the information
causality game if he and Alice have previously shared quantum as opposed to
classical correlations.\footnote{This is true in other contexts besides that of
  the information causality game. See, e.g., \citet[]{buhrman2001,
  brukner2002, brukner2004}.} And the question arises regarding the source
of this increased communicational power. But whatever that source is, it is not
the case that it manifests itself in nonlocality or nonseparability at the
\emph{operational} level.\footnote{Compare this with \citet[]{buhrman2001},
  who writes that entanglement enables one to ``\emph{circumvent} (rather
  than simulate) communication'' (p. 1831, emphasis in original), and also
  with \citet[]{bub2010}'s discussion of entanglement in the context of
  quantum computation, which he argues allows a quantum computer to compute a
  global property of a function by performing fewer, not more, computations
  than classical computers.} This is in contrast to systems whose correlations
violate the Tsirelson bound.

But the game described by \citet[]{pawlowski2009} involves the communication of
\emph{classical} bits from Alice to Bob. Might not this limitation in Bob's
ability to take advantage of his underlying connection with Alice be overcome if
we allow her to send him qubits rather than only classical bits? Indeed, it is
well known that if Alice sends a qubit to Bob that is entangled with a qubit
that is already in his possession, then Alice and Bob can implement the
`superdense coding' protocol \citep[\S{}2.3]{nielsenChuang2000}; Alice's sending
of a single qubit to Bob according to this protocol will allow him to learn two
bits' worth of classical information.\footnote{In the context of a suitably
  generalised version of the information causality game, it turns out that a
  two-bit information gain per qubit constitutes an upper bound
  \citep[]{pitaluaGarcia2013}.} Does this not undermine the claim that quantum
correlations contribute nothing over and above whatever message is sent between
Alice and Bob to the information gained by him?

It does not. On the one hand, before the transmission of the qubit(s) from Alice
to Bob, no-signalling implies that Alice and Bob can be considered as
operationally separable despite their sharing an entangled system, as we have
seen above. On the other hand, in the superdense coding protocol, after Alice
transmits her message to Bob, all of the correlated quantum system that was
initially shared is now in Bob's possession. So after transmission there is no
sense in which Bob can take advantage of correlations shared with Alice at that
time. In a sense Alice's message to Bob `just is' information regarding the
correlations that exist between them at the time at which she sends
it.\footnote{This conclusion is essentially that of
  \citet[p. 032110-20]{spekkens2007}. Fascinatingly, Spekkens also shows that
  the superdense coding protocol can be implemented in his toy classical
  theory.}

As we have seen, when Alice and Bob share PR-correlated systems, they can win a
round with certainty in the $m = 1$ game for any $N$ by exchanging a single
classical bit. Earlier I also mentioned \citeauthor[]{vanDam2013}'s
\citeyearpar[]{vanDam2013} result to the effect that PR-correlated systems allow
one to perform \emph{any} distributed computational task with only a trivial
amount of communication. These results are striking. However the reader may
nevertheless feel somewhat unimpressed by them for the following reason: the
number of PR-correlated systems required to implement these protocols, as we
have seen, is great. With respect to the length $n$ of Bob's bit string
$\mathbf{b}$ (arguably the most appropriate measure of input size for the game),
implementing the solution described above requires that they share $2^n-1$
PR-systems; i.e. the number of PR-systems required grows exponentially with the
input size. Likewise for van Dam's protocol.\footnote{Specifically, van Dam's
  \citeyearpar[]{vanDam2013} protocol requires a number of systems that can
  grow exponentially with respect to the input size of an instance of the
  Inner Product problem, after which the solution can be efficiently converted
  into a solution to any other distributed computational problem.} A reduction
in \emph{communication} complexity has therefore been achieved only at the
expense of an increase in \emph{computational} complexity. One might argue that
it is in this sense misleading to consider the complexity of implementing the
protocol with PR-correlated systems to be trivial---that they provide us with a
`free lunch'.

I will return to this point later. But for now let me say that, arguably, this
is not a relevant consideration in this context. The theories of communication
complexity and computational complexity are distinct sub-disciplines of computer
science. The goal of communication complexity is to quantify the amount of
communication necessary to implement various communicational protocols. For this
purpose one abstracts away from any consideration of how complicated a
computational system must be in other respects \citep[]{hushilevitz1997}. The
question addressed in \citet[]{vanDam2013} and in \citet[]{pawlowski2009} and
\citet{pawlowski2016} concerns whether the availability of PR-correlated systems
would make communicational, not computational, complexity theory
superfluous. From this point of view any previously prepared PR-correlated
systems are viewed as `free resources' for the purposes of the analysis.

This said, one can imagine that the subsystems of PR-correlated systems employ
some hidden means of communication with one another, and then argue that this
must be included in the complexity ascribed to the protocol. This would of
course constitute a descent below the empirically verifiable level. In itself
this is obviously not objectionable. But it is hard to see what use this would
be to a theory of communicational complexity, which after all, like
computational complexity \citep[]{cuffaro2018}, aims to be a practical science
whose goal is to guide us in making distinctions in practice between real
problems related to data transmission that are of varying levels of
difficulty. In this sense appealing to unseen and unmanipulable communication
between the subsystems of PR-systems does not help with the conclusion that
communication complexity theory, at least in an operational sense, becomes
superfluous if PR-correlated systems are available. The objection addressed in
the previous two paragraphs is nevertheless an important one that I will return
to.

Above I have motivated the idea, of \citet[p. 1101]{pawlowski2009}, that the
kind of accessibility of remote data that is possible given the existence of
correlated systems which violate the Tsirelson bound is `implausible'. I have
done so by describing, \emph{pace} \citet[]{bub2012}, the sense in which
information causality can be taken to generalise no-signalling. In so doing I
have gestured at a connection between the idea of implausible accessibility and
the \emph{prima facie} separate idea that a world in which
Tsirelson-bound-violating correlated systems exist would be `too good to be
true' in a communicational complexity-theoretic sense. My arguments have been
mainly conceptual. I have argued, that is, that a kind of conceptual ambiguity
at the operational level between the parties to a communicational protocol may
result if correlations which violate the Tsirelson bound are available to
use. As we have seen, when such stronger-than-quantum correlations are strong
enough (i.e. when $E > \sqrt{6}/3$), this results in the trivial communicational
complexity of any distributed computational task. But trivial communicational
complexity does not result, or anyway has not yet been shown to result, for
values of $E$ above the Tsirelson bounded value of $1/\sqrt 2$ that are below
$\sqrt{6}/3$. This is despite the fact that the conceptual ambiguity I have
described is present to some extent for all such values of $E$.

\begin{sloppypar}
Thus one may wonder whether `a little' ambiguity may be tolerable for practical
purposes---whether, that is, a theory which admits correlations which only
`weakly' violate the Tsirelson bound should be admitted within the space of
possible physical theories from the point of view of the information causality
principle. The situation could be seen as analogous to the situation one is
faced with in Newtonian Mechanics, in fact, for Corollary VI (which I described
in \S\ref{sec:demopoulos}) only guarantees that a system in the presence of
external forces can be treated as (quasi-) closed when these forces act
\emph{exactly} equally upon it and are \emph{exactly} parallel. Clearly this is
not the case for the Jovian system \emph{vis-\'a-vis} the sun, for
example. Corollary VI---and Proposition 3---nevertheless function as
methodological tools in that they allows us to maintain the idea of the
mutually independent existence of spatially distant things as a methodological
principle and treat the Jovian system, for the practical purpose of analysing
its internal motions, as unaffected by the forces exerted upon it by the sun.
\end{sloppypar}

There is much work to be done before information causality can be considered as
successful in ruling out---in the conceptual sense described in the previous
two paragraphs---\emph{all} theories whose correlations violate the Tsirelson
bound. Irrespective of whether this goal can be achieved, however, this does
not necessarily undermine the status of information causality motivated as a
methodological principle in something like the way that I have done in this
paper. In particular, information causality would be especially compelling if
one could draw a relation between the degree of violation of the principle and
the degree of `superfluousness' of the resulting theory of communication
complexity with an eye to distinguishing `weak' violations of the Tsirelson
bound from more objectionable violations. Thus there is much work to do in any
case.

I close with the following more fundamental objection. Why should nature care
whether beings such as us are able to engage in communication complexity
theory? In fact there is no fundamental reason why nature should
care. Analogously, there is no fundamental reason why nature should care
whether beings such as us can do physics. But the goal of empirical science is
not to derive the structure of the world or its constituent entities by way of
a priori or `self-evident' principles. It is rather to make sense of and
explain our experience of and in the world, as well as to enable us to predict
and to control aspects of that world for whatever particular practical purposes
we may have. In fact we have a science which is called physics. And in fact we
have a science which we refer to as communication complexity theory. The
principle of mutually independent existence, and analogously the principle of
information causality, may be thought of as answers to the question: `how are
such facts possible?' in the sense that they aim to identify the necessary
suppositions implicit in \emph{any} such theories and in our practice of
them.\footnote{Cf. \citet[pp. B20-B21]{kant1781german}.}

That said, these may not be definitive answers. The necessity of presupposing
Einstein's mutual independence and local action principles for the purposes of
theory testing has been questioned by \citet[]{howard1989}. In a similar way,
one might argue that it is wrong to think that the existence of correlated
systems which `strongly' violate the Tsirelson bound would make any science of
communication complexity impossible. Rather, one might conclude instead that
the idea of a science of communication complexity that is wholly independent of
\emph{computational} complexity-theoretic considerations is unachievable. This,
one might argue, is the real lesson to take away from the fact that an
exponential number of PR-correlated systems is required to implement Alice's
and Bob's solution to their guessing game. Yet even if this were all that we
learned from information causality, it would still represent a significant
advance in our understanding of the structure of our theoretical knowledge---an
understanding of the physically motivated constraints under which two
mathematical \emph{theories} may be regarded as mutually independent.

\section{Summary}
\label{sec:conc}

Above I have argued that the principle of information causality has not yet
been sufficiently motivated to play the role of a foundational principle of
nature, and I have described a way in which one might begin to provide it with
such a motivation. More specifically I described an argument, due to
Demopoulos, to the effect that the no-signalling condition can be viewed as a
generalisation, appropriate to an irreducibly statistical theory, of Einstein's
principle of mutually independent existence interpreted as a constraint on
physical practice. I then argued that information causality can in turn be
motivated as a further generalisation of no-signalling that is appropriate to a
theory of communication. I closed by describing a number of important obstacles
that are required to be overcome if the project of establishing information
causality as a foundational principle is to succeed.

\bibliographystyle{apa-good}
\bibliography{Bibliography}{}

\end{document}